Yasutomo Ota[a], Kenta Takata[b], Tomoki Ozawa[c], Alberto Amo[d], Zhetao Jia[e], Boubacar Kante[e], Masaya Notomi[b,f], Yasuhiko Arakawa[a], Satoshi Iwamoto[a,g,h]

[a] Institute for Nano Quantum Information Electronics, The University of Tokyo, 4-6-1 Komaba, Meguro-ku, Tokyo, 153-8505 Japan
[b] Nanophotonics Center and NTT Basic Research Laboratories, NTT Corporation, 3-1 Morinosato-Wakamiya, Atsugi 243-0198, Kanagawa, Japan
[c] Interdisciplinary Theoretical and Mathematical Sciences Program (iTHEMS), RIKEN, Wako, Saitama 351-0198, Japan
[d] Université de Lille, CNRS, UMR 8523 PhLAM—Physique des Lasers Atomes et Molécules, F-59000 Lille, France
[e] Department of Electrical Engineering and Computer Sciences, University of California, Berkeley, California 94720, USA
[f] Department of Physics, Tokyo Institute of Technology, H-55, Ookayama 2-12-1, Meguro 152-8550, Japan
[g] Institute of Industrial Science, The University of Tokyo, 4-6-1 Komaba, Meguro-ku, Tokyo, 153-8505 Japan
[h] Research Center for Advanced Science and Technology, The University of Tokyo, 4-6-1 Komaba, Meguro-ku, Tokyo, 153-8505 Japan

# Active topological photonics

**Abstract:** Topological photonics has emerged as a novel route to engineer the flow of light. Topologically-protected photonic edge modes, which are supported at the perimeters of topologically-nontrivial insulating bulk structures, have been of particular interest as they may enable low-loss optical waveguides immune to structural disorder. Very recently, there is a sharp rise of interest in introducing gain materials into such topological photonic structures, primarily aiming at revolutionizing semiconductor lasers with the aid of physical mechanisms existing in topological physics. Examples of remarkable realizations are topological lasers with unidirectional light output under time-reversal symmetry breaking and topologically-protected polariton and micro/nano-cavity lasers. Moreover, the introduction of gain and loss provides a fascinating playground to explore novel topological phases, which are in close relevance to non-Hermitian and parity-time symmetric quantum physics and are in general difficult to access using fermionic condensed matter systems. Here, we review the cutting-edge research on active topological photonics, in which optical gain plays a pivotal role. We discuss recent realizations of topological lasers of various kinds, together with the underlying physics explaining the emergence of topological edge modes. In such demonstrations, the optical modes of the topological lasers are determined by the dielectric structures and support lasing oscillation with the help of optical gain. We also address recent researches on topological photonic systems in which gain and loss themselves essentially influence on topological properties of the bulk systems. We believe that active topological photonics provides powerful means to advance micro/nanophotonics systems for diverse applications and topological physics itself as well.

**Keywords:** topological physics, nanophotonics, semiconductor lasers, microcavity lasers, photonic crystals, non-Hermitian photonics

# 1 Introduction

The search for novel ways to manipulate light with optical structures is a central focus of research in photonics. Topological photonics has been a key framework for such studies for recent years and keeps growing as a resource of novel concepts for advancing photonic devices, such as waveguides, beam splitters, isolators and resonators. Two dimensional (2D) photonic systems analogous to quantum Hall [1–4], quantum spin Hall [5–12] and quantum valley Hall [13–15] systems have been proposed and experimentally demonstrated in microwave and/or optical regime. Topological 1D edge states supported at the exterior of the 2D bulk systems may enable topologically-protected light transport immune to sharp waveguide bends and structural imperfections as well as unidirectional light transport inhibiting back reflection. Such robust properties, together with emerging novel functionalities, are highly attractive for building compact, low-loss and functional photonic integrated circuitry. More recently, there are attempts to revolutionize semiconductor lasers with topological photonics. Topological protection of optical modes could be useful for improving laser performances and may lead to lasers with e.g. robust single mode operation. The marriage between optical gain and topological photonic structures are also of interest for exploring novel topological physics. In this review, we overview active topological photonics that primarily explores topological photonic devices incorporating optical gain. Generally, lasers which make use of topological nature of the photonic band structure can be called topological lasers. For most cases discussed in this review, topological lasers have lasing modes from topological edge states. We focus on the introduction of various types of topological lasers reported to date and of topological phenomena that are activated by optical gain. For reviews of topological photonics itself, and its relevance to topological condensed matter physics, the readers can refer to Ref. [16–22].

Figure 1 displays a summary of different types of topological lasers. A variety of platforms have been utilized to demonstrate lasing in topological photonic structures. For supporting 0D edge states, which function as stationary cavity modes, 1D topological chains of optical resonators are frequently employed. Each site cavity can be composed of a micropillar [23] (Fig. 1(a)) or a microring resonator [24,25] (Fig. 1(b) and (c)). Replacing the site cavities with a nanoscale cavity enables the formation of topological nanocavity modes [26] (Fig. 1(d)). Another way to realize topological nanocavity is to utilize 1D topological photonic crystal nanobeams [27] (Fig. 1(e)). The interface of the two topologically-distinct nanobeams supports a tightly localized interface state. Increasing the dimensionality, 1D edge modes in 2D bulk systems can be implemented using coupled ring resonators and photonic crystals. The first report on topological lasing in a 2D structure employed a magneto-optical 2D photonic crystal that breaks time reversal symmetry (TRS), giving rise to a 1D chiral edge mode and resulting unidirectional lasing [28] (Fig. 1(f)). Lasing from helical edge states have also been examined with 2D arrays of microring resonators [29] (Fig. 1(g)); such systems can be interpreted as pseudo quantum spin Hall systems and can leverage topological protection for the propagating optical modes even without TRS breaking. Micropillar cavities are also suitable for forming 2D arrays of coupled optical resonators [30] (Fig. 1(h)).

In the following, we discuss topological lasers with different spatial dimensions and time-space symmetry. We also address novel topological phases emerging under the presence of gain and loss. In section 2, we discuss topological microcavity and microring lasers based on 0D edge states accompanied with 1D topological bulk systems. Section 3 is devoted to topological nanocavity lasers, with particular focus on the system supporting 0D edge states formed between topological photonic crystal nanobeams. In section 4, we discuss topological microring lasers based on 1D edge channels in 2D topological bulk systems with broken TRS. Section 5 reviews topological lasers implementing pseudo quantum spin Hall systems that preserve TRS. Section 6 introduces non-Hermitian topological phases enabled by gain and loss, and outlines emergent symmetry classes, together with redefined photonic bandgaps in non-Hermitian physics. Section 7 provides a summary together with a brief outlook of the field.

In this review, we focus on topology in momentum space; we do not cover active systems involving topological charge in real space [31–33]. We note that active topological photonics have also led to a variety of unique photonic systems, such as topological photonic waveguides interacting with quantum emitters [11,34] and those generating quantum light via nonlinear optical processes [35,36], and topological photonic structures subject to active temporal phase/intensity modulation [37–39] (or Floquet engineering); readers can refer to original or other review papers for these topics.

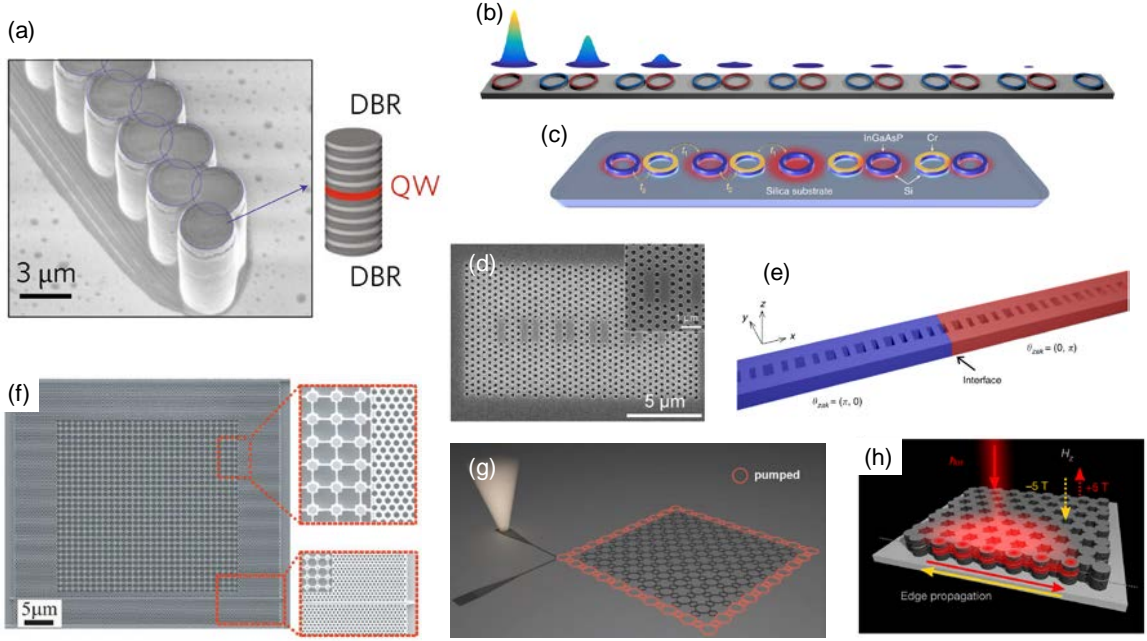

**Fig. 1:** Topological laser structures. 1D arrays of (a) micropillar cavities, (b),(c) microring cavities and (d) photonic crystal nanocavities. (e) 1D photonic crystal nanobeam with a topological disorder. (f) 2D photonic crystal forming a microring resonator with a 1D edge channel. 2D arrays of (g) microring resonators and (h) micropillars. Adapted from ref. [23] for (a), ref. [25] for (c), ref. [26] for (d), ref. [27] for (e), ref. [28] for (f), ref. [29] for (g) and ref. [30] for (h). Drawing (b) is provided by courtesy of M. Parto et al., the authors of ref. [24].

## 2 One-dimensional topological microcavity and microring lasers

The first geometry in which lasing in topological edge states was demonstrated is a one-dimensional lattice of coupled photonic resonators implementing the Su-Schrieffer-Heeger (SSH) Hamiltonian [23–25]. This tight-binding model was initially introduced to describe the electronic transport in the polyacetilene molecules and it provides one of the simplest lattice models with topological properties. The SSH lattice is schematically displayed in Fig. 2(a). Each unit cell consists of two sites with the same onsite energy (we take it as zero) forming a one-dimensional chain in which the intracell hopping $t$ is different from the intercell hopping $t'$. In momentum space, in the basis of the $A$ and $B$ sites of each unit cell of size $a$, the Hamiltonian can be written as [40]:

$$H(k) = \begin{pmatrix} 0 & t + t'e^{ika} \\ t + t'e^{-ika} & 0 \end{pmatrix}$$

The spectrum presents two symmetric bands separated by a gap of width $2|t - t'|$ [Fig. 2(b)]. The eigenfunctions of the upper (+) and lower (-) bands are [41]: $|\pm\rangle = 2^{-1/2}(e^{-i\phi(k)}, \pm 1)^\dagger$, with $\cot\phi(k) = t/(t'\sin ka) + \cot ka$. The phase and symmetry of the eigenfunctions depends on the relative value of $t$ to $t'$. The topological properties of the Hamiltonian are revealed via the winding $\mathcal{W}$ of the phase $\phi(k)$ across the Brillouin zone [41]:

$$\mathcal{W} = \frac{1}{2}\int_0^{\frac{2\pi}{a}} \frac{\partial \phi(k)}{\partial k} dk.$$

The winding $\mathcal{W}$ can have two possible values depending on the dimerization: 0 when $t > t'$, and +1 when $t < t'$. These two windings correspond to the two different topological phases of the system. In an infinite lattice, the distinction between the two cases is not relevant as it depends simply on the choice of unit cell, which defines $t$ and $t'$. However, in a semi-infinite chain, the unit cell is univocally defined by the termination, setting the winding of the chain. When $\mathcal{W} = 0$, the lattice is topologically trivial and no edge states are expected, but when $\mathcal{W} = +1$, the lattice is topologically non trivial and a topological state localised at the edge with energy in the middle of the gap (*E*=0 in our notation) is expected. Similarly, a localised interface state at *E*=0 appears when joining two semi-infinite SSH chains with different windings [40].

The topological properties of the SSH Hamiltonian are, therefore, very convenient to design structures with isolated states in the middle of the gap. In addition, these states have remarkable robustness to certain types of disorder. The topological properties of the SSH Hamiltonian are intimately related to the chiral symmetry of the Hamiltonian described above, manifested in the fact that the Hamiltonian is purely off-diagonal. The main consequence of the chiral symmetry is that the spectrum is mirror symmetric with respect to *E*=0: each eigenstate with energy *E* has a partner eigenstate at energy -*E*. The topological edge or interface state at *E*=0 is necessarily its own partner and, therefore, its energy is insensitive to any modification of the Hamiltonian that preserves chiral symmetry. In particular, disorder in the hoppings *t* and *t'* preserves this symmetry and does not affect the energy of the topological states as long as the gap between the two bands is well defined. On the contrary, the energy of the edge mode is not protected against perturbations that break chiral symmetry. For instance, disorder in the onsite energies will affect the energy of the edge mode.

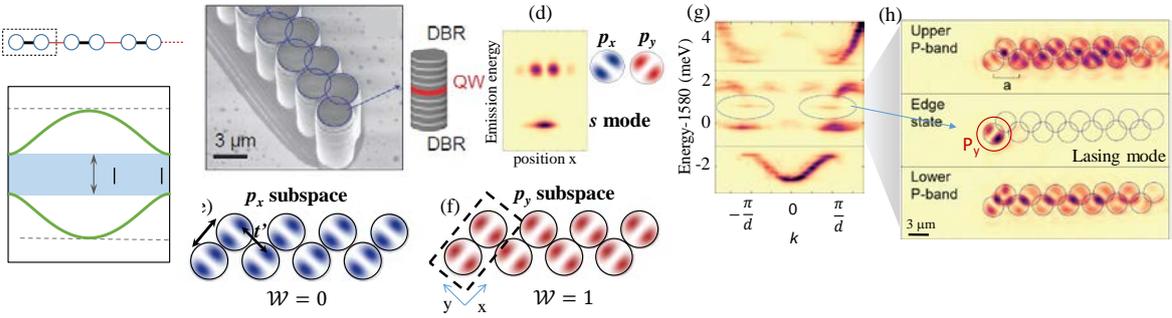

**Fig. 2:** (a) Scheme of the SSH lattice, defined by two sites per unit cell (*A* and *B*) and alternating hopping *t* and *t'*. (b) Spectrum of the eigenstates of the SSH Hamiltonian when *t*≠*t'*. (c) Realisation of the SSH Hamiltonian for *p* modes in a zigzag chain of coupled micropillars. (d) Emitted photoluminescence from a single micropillar, showing gapped *s* and *p* modes. (e)-(f) Schematic representation of the $p_x$ and $p_y$ subspaces of the zigzag chain. Each subspace implements a SSH chain with different winding. (g) Photoluminescence intensity from the zigzag chain in momentum space. It shows different bands. The middle ones correspond to the p orbitals. Emission from the topological edge state is encircled by blue ellipses. (h) Measured real-space emission at the energy of the lower and upper *p* bands, and at the energy of the edge state. The edge state has a $p_y$ geometry: this is the sub-space that contains a topological edge state. (c)-(h) adapted from Ref. [23].

In optics, the SSH Hamiltonian has been implemented making use of lattices of waveguides in fused silica [35,42,43], microwave resonators [44], and plasmonic [45] and dielectric nanostructures [46]. However, these systems are passive and cannot be used to implement lasing in edge modes. A successful strategy to circumvent this issue has been to use lattices of photonic resonators made of inorganic semiconductors, in which a quantum well layer provides a gain medium. The first system showing lasing in a topologically protected edge state was a one-dimensional lattice of coupled polariton micropillars implementing the SSH Hamiltonian [23]. In this platform, each individual micropillar possesses a series of confined photonic modes with *s*, *p*, *d*, … symmetries [Fig. 2(d)]. In Ref. [23], the SSH Hamiltonian is implemented for the $p_x$ and $p_y$ modes, whose spatial shape is strongly asymmetric. To do so, the chain of coupled micropillars was designed in a zigzag geometry [Fig. 2(c)]: due to the non-cylindrical shape of the $p_x$ modes, the alternating connecting angles between the micropillars results in mode overlap with alternating strong *t* and weak *t'* couplings of photons between adjacent sites [Fig. 2(e)]. The opposite arrangement of couplings (*t* < *t'*) is simultaneously realised for the other set of orbitals [Fig. 2(f)]. Both orbital sub-spaces present the same gap but for the chain termination shown in Fig. 2(c), only the $p_y$ subspace has a weak link at the edge and presents a topological edge state (the $p_x$ subspace has a strong link at the edge).

In order to observe lasing from the topological edge mode, St-Jean and co-workers [23] designed a microcavity with a quantum well energy such that the polariton gain was optimised for the energy of the *p* modes of the lattice. To favour gain in the topological mode vs the bulk modes, they optically pumped the chain with a spot located close to the edge.

In Refs. [24,25], the strategy was to employ the longitudinal photonic modes of evanescently coupled ring resonators made of InGaAsP, a material with optical gain at telecom wavelengths. The alternating couplings $t$ and $t'$ were engineered via the staggered separation between adjacent rings. In the work of Parto et al. (Ref. [24]), shown in Fig. 3(a)-(c), the topological state is located at the edge of the chain, where a weak coupling has been engineered. The rings were weakly coupled to guided gratings to measure the light intensity in the rings. In the work of Zhao et al. (Ref. [25]), shown in Fig. 3(d)-(e), the topological mode is located at the interface between two segments of the lattice with opposite strong/weak coupling dimerizations. In both works, to trigger lasing in the topological mode, the two groups relied on the specific shape of the topological eigenmode: due to the chiral symmetry of the SSH Hamiltonian, the eigenfunction of the topological mode at $E=0$ is nonzero only in one of the $A/B$ sublattices. In particular, it is nonzero in the sublattice corresponding to the site located at the edge of the lattice (Ref. [24]) or at the interface site (Ref. [25]). This characteristic is exclusive to that mode, all the bulk modes have components in both sublattices. Therefore, if the gain and losses of each sublattice are engineered to favour gain in the sublattice of the topological mode, lasing in this mode can be observed [47]. In Ref. [24], the authors designed an optical pumping profile such that only the sublattice of the topological mode was pumped, favouring gain in that mode. In Ref. [25], pumping was homogeneously distributed but losses were enhanced in the sublattice not containing the topological mode. This strategy had already been used to enhance the emission for the topological mode in a lattice of microwave resonators with parity-time symmetry [44].

One of the most interesting features explored in the above mentioned works is the resilience of the topological lasing mode to disorder. The chiral symmetry of the lattice Hamiltonian makes the topological mode insensitive to perturbations in the hoppings $t$ and $t'$. Perturbations in the onsite energies do affect the energy of the topological mode, which can then move up or down. However, as long as those perturbations are smaller than the bandgap, the topological mode is well isolated in the gap and lasing is preserved. This feature was experimentally demonstrated both in St-Jean et al. (Ref. [23]) and in Zhao et al. (Ref. [25]) by modifying in a controlled way the onsite energy of sites in which the topological state has a significant weight. Compared to Tamm defect modes, which emerge at the band edges when a local perturbation is added at the edge of a chain, the topological modes in a SSH lattice have the advantage of appearing directly in the middle of the gap, well separated from the bands, thus presenting a much stronger resilience to perturbations of the onsite energy. It has also been studied theoretically that topological lasing in SSH lattice is stable even in the presence of nonlinearity (saturation) in the gain strength [48].

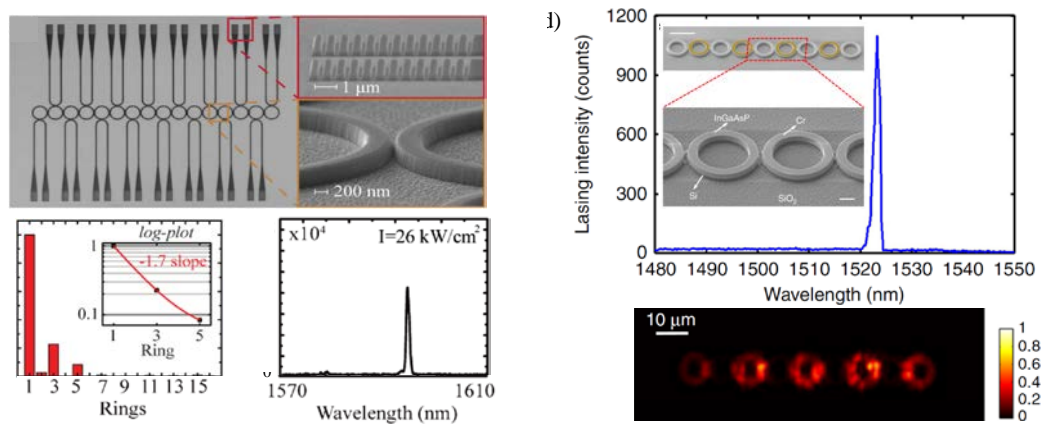

**Fig. 3:** (a) Scanning electron microscope image of the lattice of ring resonators with staggered coupling engineered by Parto et al. [24]. (b) Measured intensity of the lasing mode showing the characteristic sublattice asymmetry of the topological edge state. (c) Spectrum of the topological lasing mode. (d) Measured lasing emission from the interface state engineered by Zhao et al. [25]. The inset shows scanning electron microscope images of the coupled ring resonators fabricated on a SiO$_2$ substrate. (e) Measured spatial shape of the laser emission centred around a topological interface state. (a)-(c): adapted from Ref. [24]. (d)-(e): adapted from Ref. [25].

# 3 Topological nanocavity lasers

In the previous section, we discussed topological microcavities based on 0D edge states emerged at the perimeter of 1D bulk systems and their application to topological lasers. From the view point of future applications of topological photonics, in particular for densely-integrated nanophotonic circuits, it is imperative to consider ways to downscale such microcavities to the nanoscale for minimizing their footprints. Smaller cavities also facilitate strong light matter interactions, which improve laser performances such as spontaneous emission coupling factor, leading to low-threshold, low-power-consumption and high-speed operation. In this section, we review designs to downscale topological cavities to the nanoscale and topological lasers based on them.

A straightforward design to realize a topological nanocavity is to construct a 1D bulk system using nanoscale site resonators. With this strategy, a downsized version of the 1D SSH model discussed in the previous section can be developed. C. Han et al [26] built an array of defect-based photonic crystal nanocavities that behave as a photonic analogue of the SSH model, as shown in Fig. 1(d). The cavities are formed in an InP-based slab embedding InGaAs quantum wells, which support lasing at 0D topological edge states. The edge states were designed to tightly confine light in a resonator site and hence behaved similarly to an isolated defect nanocavity. This design of topological nanocavity results in a high Q factor over 10,000 with a small mode volume comparable to conventional high Q photonic crystal nanocavities, resulting in lasing with a high spontaneous emission coupling factor of ~ 0.1. In the same scenario, topological plasmonic nanocavities formed at the exterior of plasmonic resonator arrays can be constructed [49], though they have not been examined as laser devices.

Topological nanocavities can also be formed at the interfaces of topologically-nontrivial 1D photonic crystals, as schematically shown in Fig. 1(e). Leveraging the photonic crystal nanobeam geometry, tightly-localized nanoscale edge modes with high Q factors over 10,000 have been demonstrated [27]. Lasing from this topological nanocavity will be reviewed later in this section. Such structures do not have the chiral symmetry which was essential in the topology of SSH model. Instead, the photonic crystals hold inversion symmetry and are characterized with quantized Zak phases. A Zak phase, $\theta_{Zak}$, for an optical band of a 1D photonic crystal is defined as follows,

$$\theta_{\text{Zak}} = \int_{BZ} i \langle \psi_k | \nabla_k | \psi_k \rangle dk.$$

$|\psi_k\rangle$ is the Bloch function. $\theta_{Zak}$ is integral of Berry connection over the Brillouin zone and is equivalent to a Berry phase defined in the momentum space. A nontrivial Zak phase suggests evolution of parity of $|\psi_k\rangle$ in the momentum space. We note that one of the earliest theoretical discussion of lasing in topological edge states was reported in such a continuum photonic crystal structure, where topological edge states of one dimensional Aubry-André-Harper model was shown to operate as a laser [50].

Photonic crystal waveguides are another 1D platform to implement 0D topological nanocavities. The insertion of air holes with a period different from that of the host photonic crystal into the waveguide is known to result in a system emulating the Aubry-Andre-Harper model [51–53]. In this case, the model mimics a quantum Hall system defined in a pseudo 2D momentum space, where insertion position of the additional air holes takes the role of the momentum in an artificial dimension. Topological edge modes emerge in the virtual space as 1D edge states, which behaves as nanocavity modes in the real space with high Q factors over million.

One of other possible ideas is corner states in higher-order topological insulators [54,55], which are under intensive development also in topological photonics [56–63]. Nanocavities based on topological corner states have already been demonstrated using 2D topological photonic crystals characterized with nontrivial 2D Zak phases [63]. They can function as nanocavities embedded in 2D systems, offering a natural platform for developing integrated photonic circuits. Meanwhile, one may consider the introduction of defect-based nanocavities into 2D topological photonic crystals by, for example, fulling some airholes with dielectric [64] or introducing topological disorder [65]. These types of cavities have been far less investigated in the context of topological photonics, despite their possibilities to function as high performance nanocavities in topological nanophotonic systems.

For the rest of this section, we discuss a laser based on a topological nanocavity formed in a 1D topological photonic crystal [27], as schematically depicted in Fig. 1(e). The system under consideration is detailed in Fig. 4(a). The width and thickness of the nanobeam are chosen such that it supports only a single transverse-electric-like optical mode with a small mode cross-section. Digging square airholes in the nanobeam results in 1D photonic bandgaps, the topology of which are characterized with Zak phase defined by integral of Berry connection over the first Brillouin zone [66]. The unit cell contains two air holes located to the positions preserving inversion symmetry, which results in the quantization of the Zak phase to either 0 or $\pi$., equivalent to the quantization of the winding number $\mathcal{W}$ defined in the previous section. The relationship between the sizes of the two air holes ($d_1$, $d_2$) determines the value of Zak phase of the optical band.

Figure 4(b) shows calculated optical bands for the blue unit cell of the photonic crystal nanobeam with a period of $a$ and air hole sizes of $d_1 = 0.19a$ and $d_2 = 0.31a$ by the 2D plane wave expansion method. A wide bandgap is found between the first and second lowest energy bands. The same band structure is reproduced for the red unit cell, which possesses $d_2$ air hole at the unit cell center. One of these unit cells coincides with the other after shifting the unit cell center by a half period. This shift causes a change in Zak phase between 0 to $\pi$, hence corresponding to a change of the band topology between trivial and topological. The gapped topologically-distinct phases are connected via the point $d_1 = d_2 = 0.25a$, where the gap closes and a Dirac point is formed, as shown in Fig. 4(c). The nontrivial optical band here means the presence of a change of parity of wavefunction within the band. Figure 4(d) shows evolutions of wavefunctions for the lowest energy bands of the blue and red photonic crystals. For both cases, the wavefunctions near the zero frequency or around $\Gamma$ point behave as featureless $s$-wave. For the trivial red photonic crystal, the $s$-wave nature is preserved throughout the band. Meanwhile, for the topological case, transformation of the wave nature can be seen and an anti-symmetric wavefunction is found at X point.

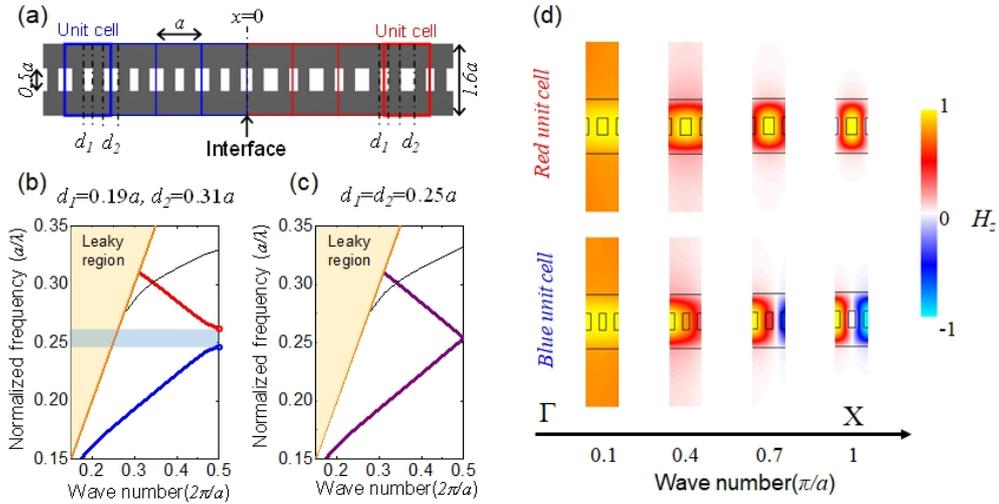

**Fig. 4:** (a) Details of the design of the topological nanocavity based on photonic crystal nanobeams, the schematic of which is depicted in Fig. 1(e). The interface is formed between a topological photonic crystal based on the blue unit cell and a trivial photonic crystal with the red unit cell. (b) Computed optical bands for the photonic crystal constituted of the blue unit cell with $d_1 = 0.19a$ and $d_2 = 0.31a$. (c) The same in (b) but with $d_1=d_2=0.25a$. (d) Evolution of wavefunctions across the momentum space, calculated for the trivial and topological photonic crystals. (a)-(c) reproduced from [27].

At the interface between the two photonic crystals with different quantized Zak phases, a single in-gap interface state is deterministically formed. This property is advantageous for photonic applications where the control of the number of modes is imperative. Let us note that contrary to the SSH lattices of photonic resonators discussed in Sec. 2, the bulk photonic crystals considered here do not hold chiral symmetry. Therefore, the energy of the in-gap state is not pinned to the zero energy [40] and the presence of the mode is guaranteed only when inversion symmetry is preserved. From a device point of view, the robustness of the topological in-gap mode to disorders in this type of photonic crystal has remained elusive, and more works are necessary to clarify it. It is noteworthy is that the emergence of the topological modes is confirmed even when the period of 1D photonic crystal becomes only a few [67]. There are several physical interpretations for the emergence of the edge mode, such as those based on the presence of edge polarization associated with nontrivial Zak phase [55,68] and on a soliton formation process via the Jackiw-Rebbi mechanism [40,65,69]. According to M. Xiao et al., the current system can be viewed as a Fabry-Pérot resonator with zero cavity length [66]. The photonic crystals function as mirrors with different reflection phases, which are related with Zak phases. A topological edge mode satisfies the resonance condition, which requires the sum of reflection phases by the two mirrors to be integer multiples of $2\pi$. This condition coincides with the case that the Zak phases of the two photonic crystals differ by $\pi$. Indeed, the current structure fulfils this condition and support one and only one in-gap mode. Figure 5(a) shows a field distribution of the topological interface mode for the design with $d_1 = 0.19a$ and $d_2 = 0.31a$, computed with the 3D finite difference time domain method. The numerically evaluated cavity Q factor is ~60,000 and the mode volume is $0.67(\lambda/n)^3$, where $\lambda$ is the resonant wavelength and $n$ is the refractive index of the host material (assuming GaAs). These values are comparable with conventional photonic crystal designs.

Experimentally, the designed topological photonic crystal nanocavity with a period of $a$ = 270 nm was formed in a GaAs slab by standard electron beam lithography and dry etching. A scanning electron micrograph of a fabricated sample is shown in Fig. 5(b). The structure embedded InAs quantum dot as gain medium. The sample was optically characterised under quasi-continuous optical pumping at a cryogenic temperature. The bottom panel of Fig. 5(c) shows a spectrum measured under a low pump power of 5 μW, exhibiting a sharp peak originated from the topological interface mode, together with broad quantum dot emission background. When increasing the pump power to 150 μm, the device underwent single mode lasing, the spectrum of which is shown in the upper panel of Fig. 5(c). A prominent cavity emission peak is observed while the background emission was strongly suppressed. From a measured light output curve, a high spontaneous emission coupling factor of 0.03 was measured, which indicates enhanced light matter interactions in the device as a result of high experimental Q factor of 9,600 and the small mode volume.

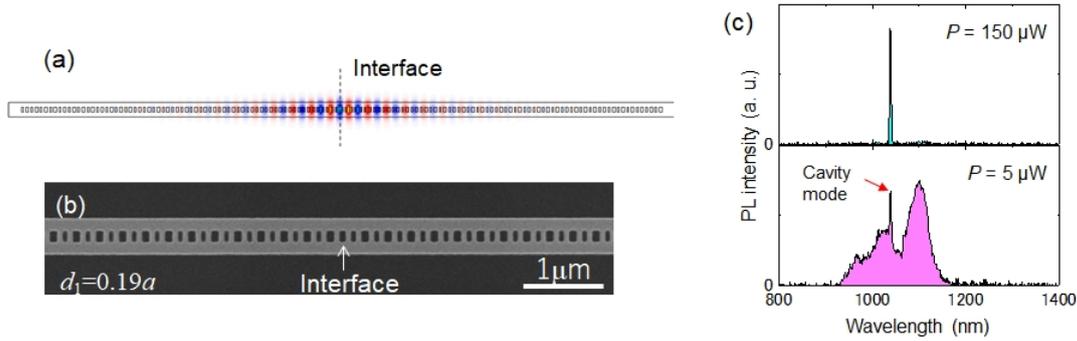

**Fig. 5:** (a) Calculated electric field distribution for the topological nanocavity designed with $d_1$ = 0.19$a$ and $d_2$ = 0.31$a$. The cavity field is well localized in the vicinity of the interface. (b) Scanning electron microscope image of the fabricated structure. The position of the interface is indicated by an arrow. (c) Emission spectra measured under optical pumping. The bottom panel shows a spectrum taken with a low pump power. The upper panel is a lasing spectrum measured with a high pump power. Clear single mode lasing can be confirmed by the single cavity peak dominating the spectrum. Reproduced from [27].

## 4  2D topological laser with TRS breaking

The possibility of opening a topological bandgap in the photonic system by breaking time-reversal symmetry was first proposed in Ref. [1] in analogy to the quantum Hall effect (QHE). The real-structure simulation was reported in Ref. [1,70] and implemented at microwave frequency [3], where the photonic crystal structure is made of ferrite rods. In this pioneering work, unidirectional transmission at a frequency within the topological bandgap is demonstrated. Such chiral edge mode is robust in a sense that backscattering is prohibited in the presence of local disorder due to time-reversal symmetry breaking of the system. The unprecedented properties of the topological edge mode bring the interest of making it as a light source, especially a laser [28,29,71]. In general, defects and disorders in the laser cavity lead to scattering loss, which degrades the performance of laser by reducing the quality factor and lowering the output efficiency. The chiral edge mode is immune to such backscattering. Besides, the topological edge mode profile depends on the shape of the topological interface, which can be designed to control the radiation pattern of the lasing mode. The lasing process, including the gain competition between the chiral edge mode and other bulk modes, is also of theoretical interest. Motivated by the above practical advantages and unsolved questions, a 2D topological laser was first built with a photonic crystal platform under an external magnetic field [28]. In the design of photonic crystal, InGaAsP multiple quantum wells are structured and bonded to yttrium iron garnet (YIG), a gyrotropic material grown on gadolinium gallium garnet (GGG). The topological band gap is opened at Γ-point under static magnetic field normal to the slab, breaking the time-reversal symmetry. The gap is designed at 1.55 μm for edge mode to lase at the telecom wavelength, which is supported by the gain spectrum of InGaAsP. To confine light at the boundary of the topological cavity, another circular unit cell embedded in a hexagonal lattice with a trivial bandgap is used (Fig. 6b).

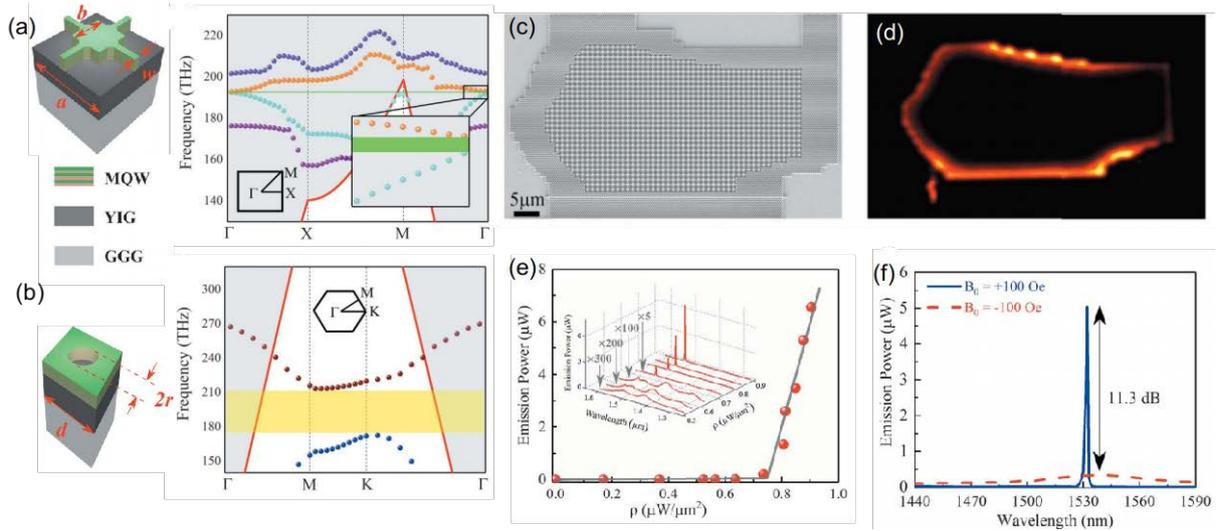

**Fig. 6:** 2D Topological photonic crystal laser with a static magnetic field. (a),(b) 3D topological (trivial) bandgap unit cell structure and corresponding band diagram. (c) SEM image of the fabricated nanostructure with topological-trivial interface in the shape of flipped US map. (d) Top lasing image of (c) under static magnetic field. (e) Emission power of the laser as a function of pumping power. (f) Output power at one end of the coupling waveguide with the opposite direction of the magnetic field normal to the structure plane. (replicated from Ref. [28])

Top view camera image shows edge mode emission at the interface when pumping the whole structure with a pulsed laser working at 1064 nm under the static magnetic field (Fig. 6d). No edge state radiation pattern is observed without magnetic bias. To further characterize the lasing behavior, a line-defect trivial waveguide is designed close to the topological boundary to couple out the power, which is collected by a lensed fiber at the end of the waveguide. A clear lasing threshold is observed with increasing pumping power. (Fig. 6c) The unidirectional property of the chiral edge mode is demonstrated by flipping the direction of the static magnetic field. In Fig. 6f, a reduction ratio of 11.3 dB in the emission power at the waveguide output proves the unidirectional propagation of the lasing mode. Different shapes of the topological-trivial interface are fabricated and similar lasing behavior is observed near telecom wavelength. We note that the topological gap of the photonic crystal under TRS breaking is narrow (tens of pm) due to limited magneto-optical response of YIG. Nevertheless, lasing from the topological edge mode was observed and further experimentally confirmed using second order intensity correlation [33]. The narrow gap compared to the free spectral range for such cavities makes their design challenging but also naturally provides a mode selection mechanism. In follow-up work, the ring-shaped topological-trivial interface is designed to generate coherent beams carrying orbital angular momentum (OAM) [33]. The device succeeds in producing arbitrarily large topological charge and multiplex different OAM emission onto a single sample. The ability to shape the edge mode profile hence the radiation pattern can be further optimized for communication applications. There are open opportunities for experimentally investigating such TRS-broken topological lasers toward practical applications.

Lasing from topological edge states in a 2D configuration has been also studied in a lattice of microcavity polaritons [30]. The building block of microcavity polariton consists of semiconductor 2D quantum wells embedded between a pair of distributed Bragg reflectors (DBR), where the excitons are strongly coupled to the cavity modes. Polaritons are thus mixed light matter quasi-particles combining the properties of its two constituents: excitons and photons. An external magnetic field perpendicular to the 2D heterostructure leads to the Zeeman splitting of excitons, and the difference in the reflection of Bragg reflector for TE and TM wave results in the effective spin-orbit coupling of photons [72,73]. The combination of both effects in a hexagonal microcavity polariton lattice (Fig. 7a) induces a topological bandgap characterized by integer band Chern numbers [74,75]. Hence the time-reversal symmetry is broken and robust chiral edge states are expected with a chirality determined by the direction of the magnetic field. Lasing in such exciton-polariton topological lattices has been reported in Ref. [30] via the mode tomography of the edge states (Fig. 7(c)). It is also worth mentioning that the gain can also be provided via electrical pumping [76].

Theoretical studies of lasing in polariton hexagonal lattices subject to an external magnetic field suggest that such states should be chiral, and significant differences between lasing in topological and trivial polariton lattices [77] should be expected. Figure 7(b) shows the band diagram of a stripe of the hexagonal lattice with finite size in x-direction but infinity in the y-direction, calculated using spinor Schrodinger equations describing the time evolution of the polariton wave function. The edge mode exists at the boundary of the lattice and the opposite group velocity indicates the unidirectional propagation. When gain is introduced in the model on one side of the edge, the eigenmode shows high amplification on that side and is expected to lase with increasing gain. The full structure of a triangular-shaped lattice with material gain along the edge is simulated for both topological (Fig. 7f) and trivial (Fig. 7g) designs. The trivial lasing mode is different from the topological one both in mode profile, as the trivial one only emits in one sublattice, and in the lasing spectrum [77].

In this section, we have seen that the 2D topological laser provides a novel type of light source. By breaking time-reversal symmetry with the magnetic field, it allows unidirectional propagating modes with arbitrary shape to emit. Interestingly, polariton lattices present significant nonlinearities [78] arising from the excitonic component of polaritons, and provide a very promising platform to study the interplay of topology and nonlinearity in the lasing regime [79,80].

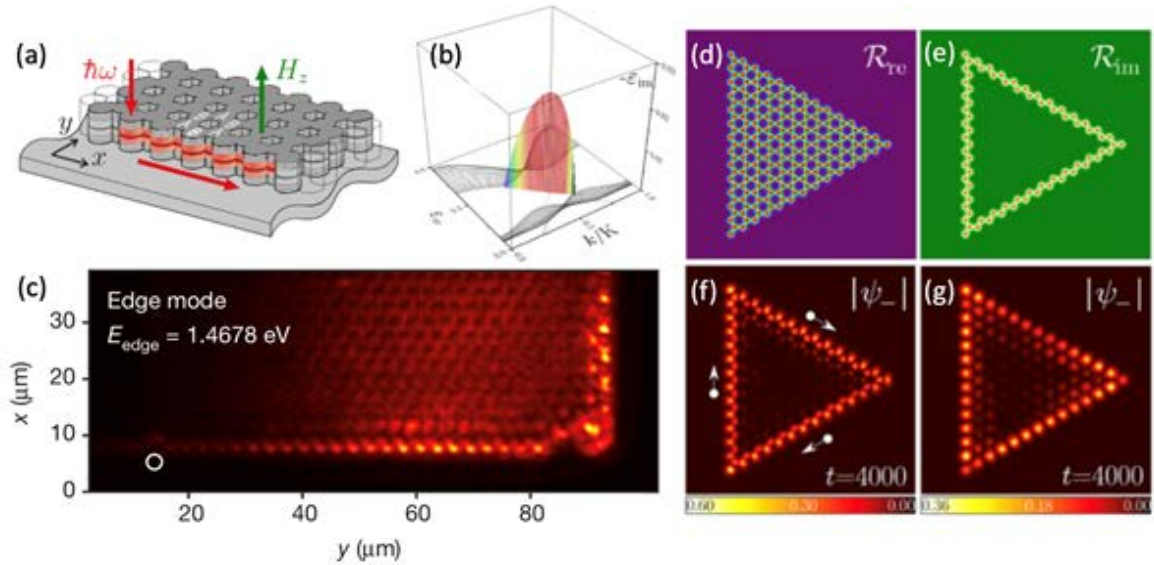

**Fig. 7:** 2D Topological polariton lasers. (a) Scheme of polariton micropillars arranged in hexagonal shape under external magnetic field. (b) Real and imaginary part of eigenmodes for a hexagonal structure with finite-y and infinite-x size. Colorbar shows that the edge mode experience amplification where $-\epsilon_{im}>0$. All $-\epsilon_{im}<0$ is set to zero in the plot. (c) Exciton-polariton topological edge state observed in photoluminescence experiment in the lasing regime. (d), (e) Real and imaginary part of the photonic potential for a triangular-shaped lattice with the gain confined at the boundary (f) Calculated unidirectional edge mode lasing profile under external magnetic field in the steady-state with arrows indicating the current direction. (g) Simulation of lasing from a trivial bandgap structure (zero spin-orbit coupling) where only sites in one sublattice emit. (Fig. (a) replicated from Ref. [74], Fig. (c) replicated from Ref [30] and Fig. (b), (d)-(g) from Ref. [77])

## 5 2D topological lasers without breaking TRS

There is another class of two-dimensional topological lasers, which do not break the time reversal symmetry [29,71]. The structure is based on the construction of the Harper-Hofstadter model using ring resonators originally realized by [5,6] in a passive silicon photonics platform.

Harper-Hofstadter model is a two-dimensional square lattice model with a uniform constant magnetic field applied perpendicular to the two-dimensional plane. In tight-binding models, the effect of the magnetic vector potential enters as complex hopping phases upon hopping to adjacent sites. The Hamiltonian of the Harper-Hofstadter model with the magnetic vector potential taken in the Landau gauge, $\vec{A} = (0, Bx)$, is

$$H(B) = -J \sum_{x,y} [a^\dagger_{x+1,y} a_{x,y} + e^{iBx} a^\dagger_{x,y+1} a_{x,y} + \text{h.c.}]$$

where we have set the lattice spacing to be unity and assumed that the hopping strength is isotropic. The phase that a particle acquires after hopping around a plaquette of the square lattice is $B$ (see Fig.8(a)). When $B = 2\pi p/q$ with $p$ and $q$ being mutually prime integers, the model has $q$ bands, and all the bands are topologically nontrivial unless $p/q$ is an integer or a half-integer. The topological invariants describing the system are Chern numbers, which equal the number of modes circulating around the system due to the bulk-edge correspondence. Hafezi et al. [5,6] realized the Harper-Hofstadter model by connecting Si microring resonators via anti-resonant link resonators which are constructed so that the path length for photons hopping from one resonator to another is generally different from the path length hopping in the opposite direction. This path difference effectively introduces the complex hopping phase necessary to construct the Harper-Hofstadter model (see Fig.8(b)).

Harper-Hofstadter model itself breaks the time-reversal symmetry unless $B$ is an integer or a half-integer multiple of $2\pi$. In the construction of [5,6], time reversal symmetry was indeed not broken (there is no external magnetic field). In fact, the Si microresonator support two degenerate modes: in one of them the light circulates clockwise mode and in the other it circulates counter-clockwise mode. The clockwise mode obeys the Harper-Hofstadter model with magnetic field strength $B$, while the counter-clockwise mode obeys the Harper-Hofstadter model with opposite magnetic field $-B$. The time reversal symmetry of the full system is preserved, and the overall Hamiltonian in the basis of clockwise and counter-clockwise modes, which serve as pseudospins, is then:

$$H_{\text{overall}} = \begin{pmatrix} H(B) & 0 \\ 0 & H(-B) \end{pmatrix}$$

This Hamiltonian possesses two time-reversal symmetries (anti-unitary operators which commute with the Hamiltonian), $T_1 = \sigma_x K$ and $T_2 = \sigma_y K$, where $K$ is complex conjugation and the Pauli matrices act on the pseudospin basis. Note that $KH(B)K = H(-B)$. There is also an on-site pseudospin conservation symmetry $U = \sigma_z$, which is essentially the product of $T_1$ and $T_2$. As long as $U$ is unbroken, and thus there is no coupling between clockwise and counter-clockwise modes, the system behaves as two copies of the Harper-Hofstadter models with opposite magnetic fields. The system is thus similar to the time-reversal symmetric quantum spin-Hall Hamiltonian in electronic systems [81–83]. However, it is worth mentioning the difference between the time-reversal symmetric quantum spin-Hall systems of electrons and its photonic analog. The model with unbroken $U$ possesses both bosonic and fermionic time reversal symmetries, because $T_1^2 = +1$ and $T_2^2 = -1$. In electronic quantum spin-Hall systems, one can allow perturbations mixing two spins which break $U$ and $T_1$ but not $T_2$, and the resulting model is described by the $Z_2$ topological invariant characteristic of two-dimensional fermionic systems with time-reversal symmetry [81]. On the other hand, in bosonic systems such as photonics, time-reversal invariant perturbations will in general mix two pseudospins and break $U$ and $T_2$ but not $T_1$, which renders the symmetry class of the model AI in the ten-fold way classification of topological insulators and superconductors, which is topologically trivial in two dimensions [84,85]. This implies that topological features of the model are protected as long as the pseudospin conservation is obeyed. The experiment of [5,6] observed negligible backscattering and hence the system is well described as two decoupled copies of the Harper-Hofstadter models with opposite magnetic fields.

The authors of [29] used ring resonators made of InGaAsP, a material with optical gain, and constructed the Harper-Hofstadter model with $B = 2\pi/4$. Resonators are made of waveguides which operate at wavelength of 1550nm. See Fig.8(c) for the images of the experiment. To obtain lasing at topological edge states, the authors of [29] added gain only to the perimeter of the system; in this way, the overlap of the gain profile with the topological edge state becomes much larger than that with the bulk states, and the topological lasing is preferred [71]. Fig.8(d) shows the comparison of lasing between the topological Harper-Hofstadter model with $B = 2\pi/4$ and a simple square lattice, which is topologically trivial. For both topologically trivial and nontrivial cases, only the perimeter of the system is pumped. The authors found that the lasing is much more efficient for the topological case. Emission spectrum has only one peak for topological case, whereas the trivial case has a much broader emission spectrum. The broad emission spectrum for the trivial case is attributed to fragmented lasing at several different segments of the perimeter, where localized modes exist due to intrinsic disorders present in the system. On the other hand, the lasing of topological case occurs from topological edge states, which are extended around the entire perimeter of the system and thus no fragmentation of lasing is observed. The authors of [29] further studied the propagating nature of lasing from the topological edge states. By pumping only a part of the perimeter of the system and observing the emission at output couplers situated at a corner of the system, they observed that lasing from topological edge states propagate and reach the output coupler much more than lasing from the perimeter of the trivial model. The topological lasing from propagating topological edge states has advantages that the entire edge can participate in lasing from a single mode due to the extended nature of the edge states. Such single-mode topological lasing is robust even against disorders at edges. Single-mode lasing and robustness against disorders are confirmed in the experiment of [29]. Nevertheless, further and careful investigations are necessary to clarify the potential of practical use of this type of topological lasers. One concern is its narrow topological bandgap (less than 1 nm), which is comparable to that of the laser linewidth (full width) well above the threshold. The stability of such laser device under harsh environments has not been investigated.

Lasing in propagating topological edge states of the Harper-Hofstadter model has also been theoretically studied in [86,87]. The authors of [86] found that pumping only a part of the edge has higher lasing threshold compared to the case where the entire edge is pumped. When only a part of the edge is pumped, even when exponential growth of an initial perturbation starts to happen, the amplified region travels away from the pumping region due to nonzero group velocity of chiral edge states, and the amplitudes decay back to zero; such form of instability is called *convective instability*. In order to achieve *absolute instability* and stable lasing, higher lasing threshold is required. The authors of [86] also found that topological lasing from pumping only a part of the edge is much less robust against disorders compared to the case where the entire edge is pumped.

When the time reversal symmetry is not broken, there are two degenerate topological edge states, one made of clockwise modes and the other made of counter-clockwise modes; these constitute helical edge states of the system. In the experiment, lasing can therefore occur from both helical edge states. To break the time-reversal symmetry, the authors of [29] also constructed the Harper-Hofstadter lattice made of S-shaped resonators (see Fig.8(e)). The spatial asymmetry of the S-shape combined with nonlinearity/loss leads to lasing in a particular mode, achieving lasing in topological edge states propagating only in one direction (see Fig.8(f)).

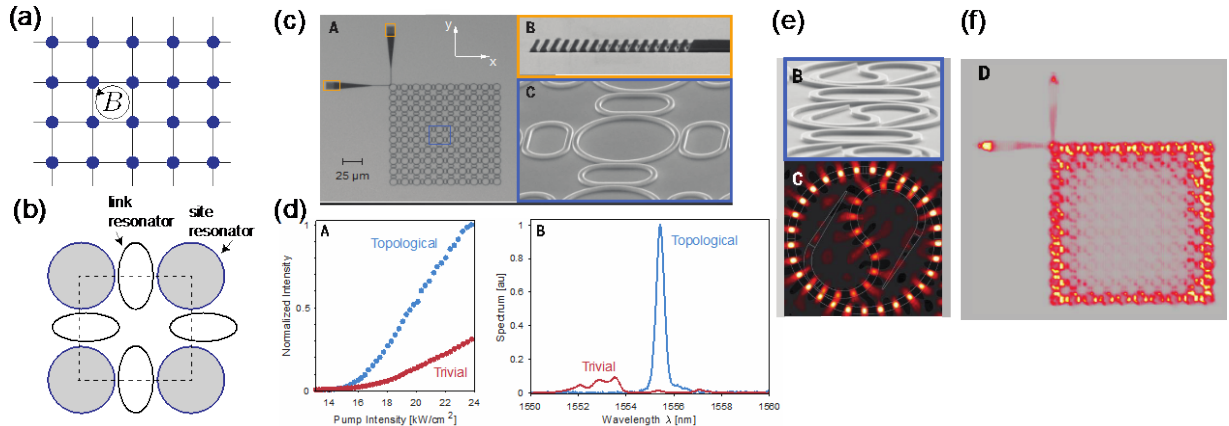

**Fig. 8:** (a) A schematic picture of the Harper-Hofstadter model. A particle hopping around a plaquette of the lattice acquires a phase *B*. (b) How to construct a plaquette of the Harper-Hofstadter model with resonators. Site resonators are connected by link resonators which are anti-resonant with modes of the site resonators. Link resonators connecting site resonators horizontally are positioned symmetrically respect to hopping to right and left, whereas link resonators connecting vertical sites are asymmetrically positioned. This asymmetry introduces different optical paths for photons going up with respect to those going down, resulting in an effective hopping phase upon vertical hopping. (c) Image of the experiment of [29]. The Harper-Hofstadter model with 10 x 10 lattice sites are constructed. There are output couplers at the upper-left corner. (d) Intensity of the emission when the perimeter of the system is pumped as a function of (A) pump intensity and (B) wavelength. (e) Resonators with S-shaped waveguides. The lower figure is the numerically calculated field distribution of a resonator with an S-shaped element. (f) Experimentally measured intensity profile of the lasing edge mode from the topological Harper-Hofstadter model made of resonators with S-shaped elements. The intensity difference in the output couplers indicate that the chiral edge state propagating in the counter-clockwise direction has much higher intensity than that propagating in the opposite direction. (c) – (f): adapted from Ref. [29].

# 6 Non-Hermiticity-based photonic topological effects

Because practical optical cavities (waveguides) are open dissipative systems and lasers (modulators) utilize amplification (absorption), photonic devices can be understood as a semi-classical analogy of non-Hermitian quantum systems [88] — they do not preserve the number of photons. Uniform optical gain or loss just adds an offset imaginary term to the eigen-spectrum for the original Hermitian system. Thus, topological lasers with homogeneous or gently distributed pumping essentially take over the topological properties of their underlying Hermitian structure. Many of the aforementioned examples in this article fall in to this category of *pumped Hermitian topological systems*.

On the other hand, specific non-Hermiticity, e.g. alternating gain and loss and asymmetric couplings, has recently been found to trigger unconventional topological functionalities, concepts and phenomena, such as reconfigurable topological insulating phases by gain and loss, extended classification of symmetry-protected topological phases, new types of bandgaps and emergent topological features, and modified bulk-boundary correspondence, etc. Extensive theoretical investigations on these topics have been proceeding, and photonic technology will play a pivotal role in opening up this new arena, as it has done in the field of non-Hermitian or parity-time- (PT-) symmetric optics [89–93].

This section briefly overviews the recent progress in non-Hermitian topological physics and photonics, so that it hopefully enhances photonics research into this exciting direction.

## 6.1 The complex SSH model

Studies on the impact of non-Hermiticity on topological photonics have branched off from the trend of exploring PT-symmetric photonic devices. The seminal work by Schomerus [47] studied the one-dimensional photonic SSH model [94] with staggered on-site gain and loss, called complex SSH model [Fig. 9 (a)],

$$H_{cSSH}(k) = \begin{pmatrix} -i\gamma & \eta + \kappa e^{-ika} \\ \eta + \kappa e^{ika} & i\gamma \end{pmatrix},$$

where $k$ is the Bloch wavenumber, $\gamma$ the magnitude of gain ($i\gamma$) and loss ($-i\gamma$) rate, $a$ the spatial period of the unit cells, and $\eta$ and $\kappa \in \mathbb{R}$ are the distinct cavity couplings determined by their alternating spatial intervals. Here, it was shown theoretically that a topological defect state was feasible [Fig. 9 (b), (c)], regardless of the following peculiar features of the system. First, the topological edge and defect states have finite imaginary eigenvalues [Fig. 9 (b)], namely net gain or loss in their dynamics [95,96]. Second, the conventional Zak phase [97] becomes continuous then fails to be a good topological number of this non-Hermitian system [98,99]. Moreover, as seen in the system eigen-detuning, $\Delta\omega_{cSSH}(k) = \pm\sqrt{\kappa^2 + \eta^2 + 2\eta\kappa\cos(ka) - \gamma^2}$, the bandgap in the real part of frequency (frequency bandgap) closes for $|\gamma| > |\kappa - \eta|$ with an exceptional point (EP), i.e. non-Hermitian degeneracy at a defective point [Fig. 9 (d)]. Here, the bulk eigenstates crossing an EP undergo a transition from distributed modes to localized modes (PT symmetry breaking). With a finite frequency gap remaining in the system, the relevant topological states are experimentally observable, even though they have finite net gain or loss [42,44,100].

In the complex SSH system with open boundaries, the topological edge states undergo the PT symmetry breaking independent of that of the bulk modes. Thus, as shown in Sec. 2, one of the robust edge states can be configured to be the only mode that obtains net gain, suppressing the mode competition in lasing [24,25] (Fig. 3). Moreover, although small finite SSH lattices exhibit non-negligible frequency (propagation constant) splitting between the weakly coupled edge states, the gain and loss can cancel it and give the exact zero modes at the resultant EP [101].

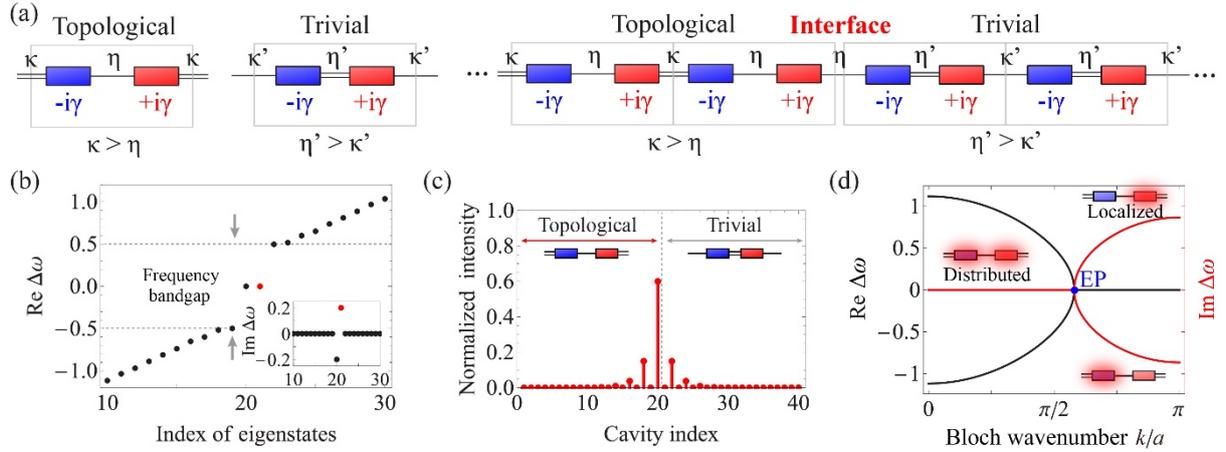

**Fig.9:** The complex SSH model [47], where staggered imaginary potential is applied to the system of periodic optical dimers based on two alternating couplings κ and η. Here, κ and η are determined by the system structure, i.e. distinct spatial intervals between the cavities. (a) Left: the topologically nontrivial and trivial unit cells. Right: A junction between the topological and trivial lattices with opposite magnitude relations of the couplings. The system forms a robust topological defect state on the interface. (b) Eigen-frequency spectrum Δω of the system with the junction shown in (a) and 20 cavities for each of the topological and trivial parts. κ = η' = 1, η = κ'= 0.5, γ = 0.2. The red dots show the complex eigen-frequency of the topological defect state. The other zero mode is the edge state of the topological lattice. (c) Intensity distribution of the topological defect state. (d) Band structure with large gain and loss. κ = 1, η = 0.5, γ = 1. The gain and loss makes the bandgap close and form an EP. The eigenmodes with imaginary detuning beyond the EP are localized at either of the cavity with gain or that with loss.

## 6.2 Gain- and loss-induced photonic topological insulating phase

Particular interest in non-Hermitian topological photonics is on new functionalities and phenomena that have no counterparts in Hermitian systems. From the technical viewpoint, it is common to control effective on-site imaginary potential in optical devices, i.e. gain and loss in lasers, amplifiers and modulators. Such an operation can also be fast, effective and local compared to the handling of the real part of the refractive index, which needs intrinsically small optical nonlinear or electro-optic effects. However, Hermitian topological optical devices rely mostly on their built-in structures and magneto-optical effects for their nontrivial photonic topology [16–18,21]. Thus, it is a technical challenge to achieve fine-grained control of the topology and resultant features such as the number and position of robust topological edge states, after the devices are fabricated. If there is a scheme to manipulate photonic topology only with gain and loss (via pumping), topological photonic modes, which are robust to structural fluctuation inevitable in tiny devices, can be incorporated in a suite of element technology for reconfigurable optical circuits.

A simple and unique scheme to enable the idea in one-dimensional coupled cavities (and waveguides) has been proposed in Ref. [102]. In the complex SSH model initially studied, the difference of imaginary potential in the unit cell can close but not open the frequency bandgap. Thus, it can break but not produce the topological insulating phase. In contrast, the proposed method [102] prepares a linear cavity array with uniform couplings (κ) as a system with identical four-cavity periods [Fig. 10 (a)]. Here, gain is introduced to two sequential cavities; loss is applied to the other two, and they are balanced under two magnitudes ($g_1$ and $g_2$) so that the system's effective Bloch Hamiltonian is given by,

$$H_{TN}(k) = \begin{pmatrix} ig_1 & \kappa & 0 & \kappa e^{-ika} \\ \kappa & -ig_2 & \kappa & 0 \\ 0 & \kappa & -ig_1 & \kappa \\ \kappa e^{ika} & 0 & \kappa & ig_2 \end{pmatrix}.$$

In this condition, gain and loss can induce the frequency bandgap and even a transition between topological and trivial photonic insulating phases.

Effective dimerization by the cavities with gain and loss can be understood as the physical mechanism of the peculiar topological property. As mentioned above, the contrast of on-site imaginary potential can cancel the splitting of the eigenvalues and result in the transition to the localization of eigenstates [103]. In the system of the four-cavity periods [Fig. 10 (a)], while the couplings between the cavities with gain and those with loss are effectively suppressed [Fig. 10 (b)], both the two cavities with gain and those with loss remain bound. As a result, despite that the Hermitian system without gain or loss ($g_1 = g_2 = 0$) gives gapless dispersion [Fig. 10 (c)], the non-Hermitian unit cell ($g_1, g_2 \neq 0$) behaves as a pair of dimers, which leads to the frequency bandgap [Fig. 10 (d)]. In addition, the array exhibits the topological insulating phase with a couple of midgap edge states, if the cavities with gain (or loss) are located separately at the both ends of the unit cell ($g_1 g_2 > 0$). Otherwise, the system is in the topologically trivial phase that results in no topological state ($g_1 g_2 < 0$). The topological transition of this single-gap non-Hermitian system can be confirmed by the discrete change in the non-Abelian global Berry phase (total complex Berry phase) with the biorthogonal basis [104]. The photonic topology is based on a non-Hermitian symmetry called pseudo-anti-Hermiticity, $\Gamma H_{TN}^\dagger(k) \Gamma^{-1} = -H_{TN}(k)$, where $\Gamma = \Gamma^{-1} = I_2 \otimes \sigma_z = \text{diag}(1,-1,1,-1)$ here [105,106]. Thus, as long as the cavity coupling is reciprocal, Re $\Delta\omega = 0$ of the edge states is protected against fluctuation of all the parameters in the model, $\kappa$, $g_1$ and $g_2$. Topological winding of the bulk coupling parameters of $H_{TN}(k)$ around Re $\Delta\omega = 0$ can also be visualized.

The gain- and loss-based topological array has a pair of topological edge states, each of which localizes at each edge due to parity symmetry breaking [Fig. 10 (e)]. In addition, by the electrically or optically tuned pumping for every cavity, which is feasible in on-chip laser systems [23–25,28,32,107,108], it is possible to introduce both topological and trivial insulating parts with their sizes and position manipulated freely in a single array. As a result, topological defect states, formed at every interface between the two distinct parts, are controlled accordingly [Fig. 10 (f)]. It is noteworthy that the reconfigurable photonic topology can be temporally handled and compatible with recent approaches that embed wave topology in the system dynamics itself [109,110]. The scheme has also been extended to a second (higher) order non-Hermitian topological insulator, which exhibits degenerate topological corner states in a two-dimensional system [111]. It would further inspire the gain- and loss-based topology for two- and three-dimensional non-Hermitian systems, which is already been under intense research [112–120]. A systematic technique to construct topological tight-binding models [121] might be extended and applied to non-Hermitian Hamiltonians.

It can be emphasized that the band engineering with gain and loss here offers a new avenue for accessing to a non-Hermitian topological insulating phase controlled solely by non-Hermitian degrees of freedom, in contrast to the complex SSH model.

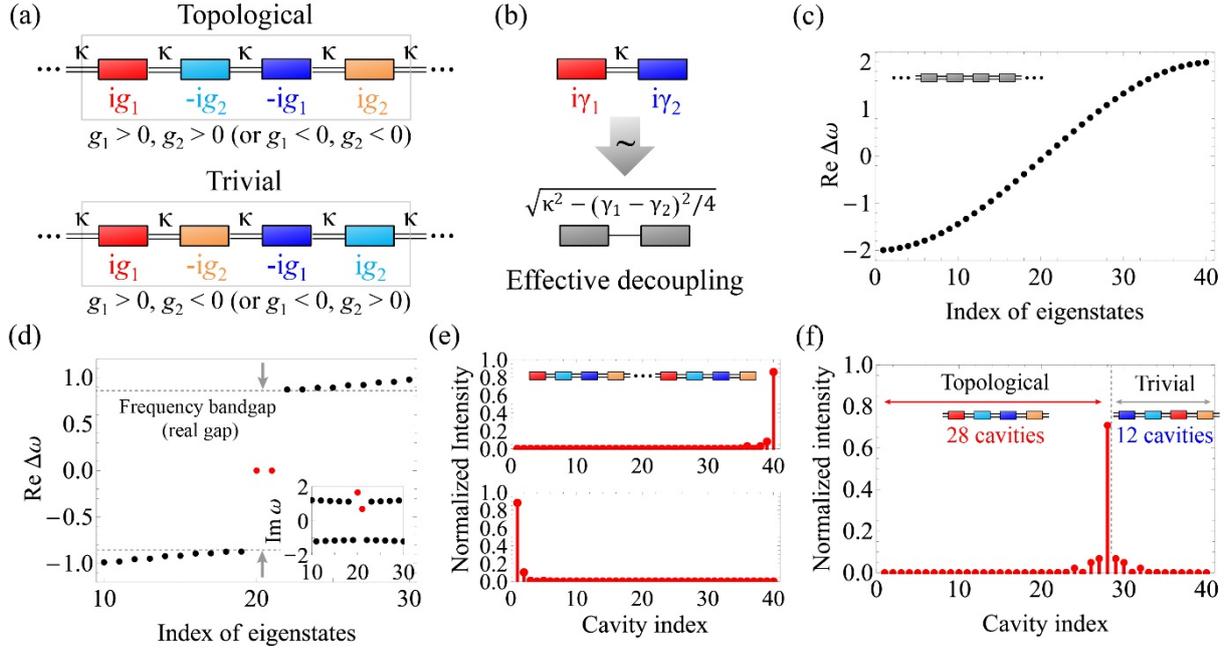

**Fig.10:** Gain- and loss-induced one-dimensional photonic topological insulating phase [102]. (a) Four-cavity unit cell of the model. Top and bottom correspond to topologically nontrivial and trivial systems, respectively. (b) Reduction of frequency splitting of supermodes by the contrast of imaginary potential. It works as effective decoupling that is especially enhanced when the signed magnitudes $\gamma_1$ and $\gamma_2$ have opposite signs (i.e. the cavities have gain and loss). (c) Eigen-detuning of the system of 40 cavities with open ends and no gain or loss. It traces the system band structure that is gapless and cosinusoidal. $\kappa = 1$. (d) Complex eigen-detuning profile of the same lattice with the gain and loss introduced, so that the system becomes topologically nontrivial. It has a bandgap in Re $\Delta\omega$ (real gap), and the red dots correspond to the topological edge states. $\kappa = 1$, $g_1 = 2$, $g_2 = 1$. (e) Intensity distributions of the edge states in (d). Each of them localizes at each edge of the lattice. (f) A topological defect state formed at the interface between topological and trivial parts in a single cavity array. The photonic topology of them depends solely on the gain and loss, thus the position and number of such topological states can be controlled by the pumping profile of the system.

## 6.3 Classification of symmetry-protected non-Hermitian topological phases

Recently, systematic classifications of symmetry-protected non-Hermitian topological systems have appeared [122–124]. Among the literature, a consistent framework by Kawabata *et al.* [123] points out that, due to the inequality of the transpose and complex conjugate of non-Hermitian single-particle Hamiltonian, $H^T \neq H^*$, some symmetries that have been regarded equivalent ramify, and some thought distinct are unified as well [116]. In short, Ref. [123] redefines Altland-Zirnbauer (AZ) symmetry classes [125] as ones comprising the conventional time-reversal symmetry (TRS) and particle-hole symmetry (PHS), which those of second-quantized systems originally reduce to,

$$T_+ H^*(\boldsymbol{k}) T_+^{-1} = H(-\boldsymbol{k}), \quad T_+ T_+^* = \pm 1 \quad \text{(TRS)},$$
$$C_- H^T(\boldsymbol{k}) C_-^{-1} = -H(-\boldsymbol{k}), \quad C_- C_-^* = \pm 1 \quad \text{(PRS)},$$

where $T_+$ and $C_-$ are unitary matrices. On the other hand, their dual counterparts with $H^T$ and $H^*$ switched should be treated distinctly as *variant* symmetries, denoted by TRS$^\dagger$ and PRS$^\dagger$,

$$C_+ H^T(\boldsymbol{k}) C_+^{-1} = H(-\boldsymbol{k}), \quad C_+ C_+^* = \pm 1 \quad (\text{TRS}^\dagger),$$
$$T_- H^*(\boldsymbol{k}) T_-^{-1} = -H(-\boldsymbol{k}), \quad T_- T_-^* = \pm 1 \quad (\text{PRS}^\dagger),$$

and they form another set of symmetry classes named AZ$^\dagger$ symmetry. Here, non-Hermitian chiral symmetry (NH CS), based on the combination of each pair of symmetries, is described in the common form as,

$$\Gamma H^\dagger(\boldsymbol{k}) \Gamma^{-1} = -H(\boldsymbol{k}), \quad \Gamma^2 = 1 \quad (\text{NH CS}).$$

Finally, sub-lattice symmetry (SLS) hence gets separate from CS as

$$S H(\boldsymbol{k}) S^{-1} = -H(\boldsymbol{k}), \quad S^2 = 1 \quad (\text{SLS}).$$

With all of them considered, the classification becomes 38-fold in non-Hermitian physics, indicating significant extension from the celebrated 10-fold way [84,126] for Hermitian topological insulators and superconductors [127,128]. Note that NH CS is equivalent to pseudo-anti Hermiticity. Remarkably, it is distinct from the prevalent form of (Hermitian) chiral symmetry, $\Gamma H(\boldsymbol{k}) \Gamma^{-1} = -H(\boldsymbol{k})$.

Photonic systems with imaginary on-site potential all satisfy $H^T \neq H^*$, and they hence break TRS in most cases. Thus, it is reasonable to start with checking the list of A, AIII and AZ$^\dagger$ symmetry classes for classifying such systems. For example, the complex SSH model is in AIII class of the non-Hermitian version (Complex AIII) characterized by NH CS [123]. The abovementioned four-cavity model $H_{TN}(k)$ [102] actually respects TRS$^\dagger$ with $C_+ = I_4 = \text{diag}(1,1,1,1)$, PRS$^\dagger$ with $T_- = \text{diag}(1,-1,1,-1)$, and hence NH CS with $\Gamma = T_-$. Thus, in terms of symmetry, it is classified into an unconventional class named BDI$^\dagger$. It can nonetheless be interpreted as Complex AIII due to the criticality of NH CS on the topology as well, and the two classes are equivalent in one dimension. The theory [123] tells that when the system has a "real gap," meaning exactly the same as frequency bandgap here, it is characterized by an integer ($\mathbb{Z}$) index. This supports the correspondence between the edge states and global Berry phase of the non-Hermiticity-based topologically insulating lattice. Applying the encompassing table of non-Hermitian symmetries will promote further the understanding of non-Hermitian topological insulating systems and help with systematic design of them.

Here, PT symmetry in the bulk complex SSH system, i.e. $\sigma_x H_{cSSH}^*(k) \sigma_x = H_{cSSH}(k)$, does not affect the topological invariant and edge states under the frequency gap, provided by NH CS. Instead, it protects the system from the "non-Hermitian skin effect" mentioned later, where the bulk modes concentrate over an edge via non-Hermiticity [123].

## 6.4 Complex bandgap and emergent non-Hermitian topological effects

Another important fact in non-Hermitian physics is that it is necessary to redefine the *bandgap* because eigen-spectra of non-Hermitian systems are generally complex [113,123,129]. One of the two possible types of "complex bandgaps" is called a *line gap* and defined as a line-shaped blank of eigenstates in the bulk spectrum plotted on the complex plane [Fig. 11 (a)]. The other is termed a *point gap*, which means a (symmetry-protected) frequency point that the spectrum does not cross (except for the case of no symmetry) [Fig. 11 (b)]. Line gaps are divided further into two, and one is the forementioned real gap indicating the discontinuity in the real part of eigen-frequencies (i.e. the line gap parallel to the imaginary axis). The other is absence of states for a range of their imaginary part, called an imaginary gap. A remarkable point is that possible system topological indices also depend on which complex gap the system possesses [123].

A topological structure specific to non-Hermitian systems shows up, especially when their complex spectra form loops around point gaps [129]. In this case, the system is characterized by the geometric charge not of eigenstates but rather of the band spectrum $H(\mathbf{k})$ for a certain loop in the $k$ space: $w_n = 1/(2\pi) \oint dk \, \partial_k \arg H_n(\mathbf{k})$, where $n$ is the eigenstate index and $H_n(\mathbf{k})$ is the $n$th eigenfrequency. The non-Hermitian winding number $w$ here is considered as the sum or difference of $\{w_n\}$ among distinct states and called vorticity [112,113,129]. In one dimension, at least, systems with point gaps and finite $w$ are closely related to lattices with asymmetric hopping [Fig. 11 (c)]. The pioneering work reports that such a system can have edge states, but they can be dynamical and unstable [129].

It is also known that non-Hermiticity can explicitly disrupt the conventional bulk-boundary correspondence (BBC) held in Hermitian systems [119,130,139–141,131–138]. Especially, asymmetric couplings make not only topological edge states but also non-topological bulk states localize around an either end to which the stronger hopping is directed (non-Hermitian skin effect) [Fig. 11 (d)]. Regarding *line-gapped* systems, the topological transition in this case is explained by the winding number of the $Q$ matrix based on the biorthogonal projection operator [132,133,140]. Furthermore, imaginary gauge fields [135,142–144], which indicate paired effective amplification in one way and attenuation to the other, are presented as an intelligible interpretation of these anomalous properties. The skin effect will hence stem from some symmetry breaking induced by non-Hermiticity [135,145]. General complex asymmetric (directional) couplings can break inversion symmetry and all the elemental symmetries (TRS$^{(\dagger)}$ and PRS$^{(\dagger)}$) for the AZ and AZ$^\dagger$ classes. A latest study [146] points that the skin effect is unavoidable in *point-gap* topology. Moreover, once it occurs in any systems with open boundaries, covering most of its experimental realizations, it inevitably closes the bulk point gap and washes out the relevant topological protection of the boundary modes. The work shows a new way to circumvent the collapse of BBC via a peculiar skin effect retaining TRS$^\dagger$.

For the purely one-dimensional optical lattice comprising single-mode elements, the directional hopping means non-reciprocity, and the foregoing anomaly is hence considered unrealizable as a static response of standard dielectric devices. Meanwhile, ring cavity arrays with ancillary components are proposed to induce imaginary gauge fields in supermodes with a specific chirality [Fig. 11 (e)] [112,143,144].

Finally, it is noteworthy that exceptional points (EPs) [91–93,147–153], i.e. defective complex-gap closings, will be classified separately [154]. They can also be symmetry protected [155–157] and are featured by topological indices that can lead to topological states. The topological classification of EPs suggests that some robust photonic defect states in early reports would be EP-based [112,158,159]. With a certain range of parameters, the four-cavity model $H_{TN}(k)$ can also have a couple of EPs (point-gap closings) on Re $\Delta\omega = 0$, which are protected by NH CS and separated by the imaginary gap. These EPs have distinct fractional charge vortices $w_n = \pm 1/2$ and can hence lead to edge states in the system with the real gap closed. Technically, however, some gaps are still to be filled for the comprehensive understanding of the BBC in systems with EPs. On the experimental side, (right) eigenstates of non-Hermitian systems are generally non-orthogonal, and photonic states with the same real frequency, which especially centre at EPs, can hence couple to each other. Thus, cares might be taken to selectively control topological states in non-Hermitian optical systems. Meanwhile, recent judicious demonstration [24,25,42,44,100,160,161], which has been reaching controlled light steering in a two-dimensional lattice [162], suggests a bright outlook on applying the series of novel non-Hermitian topological properties for future photonic circuits.

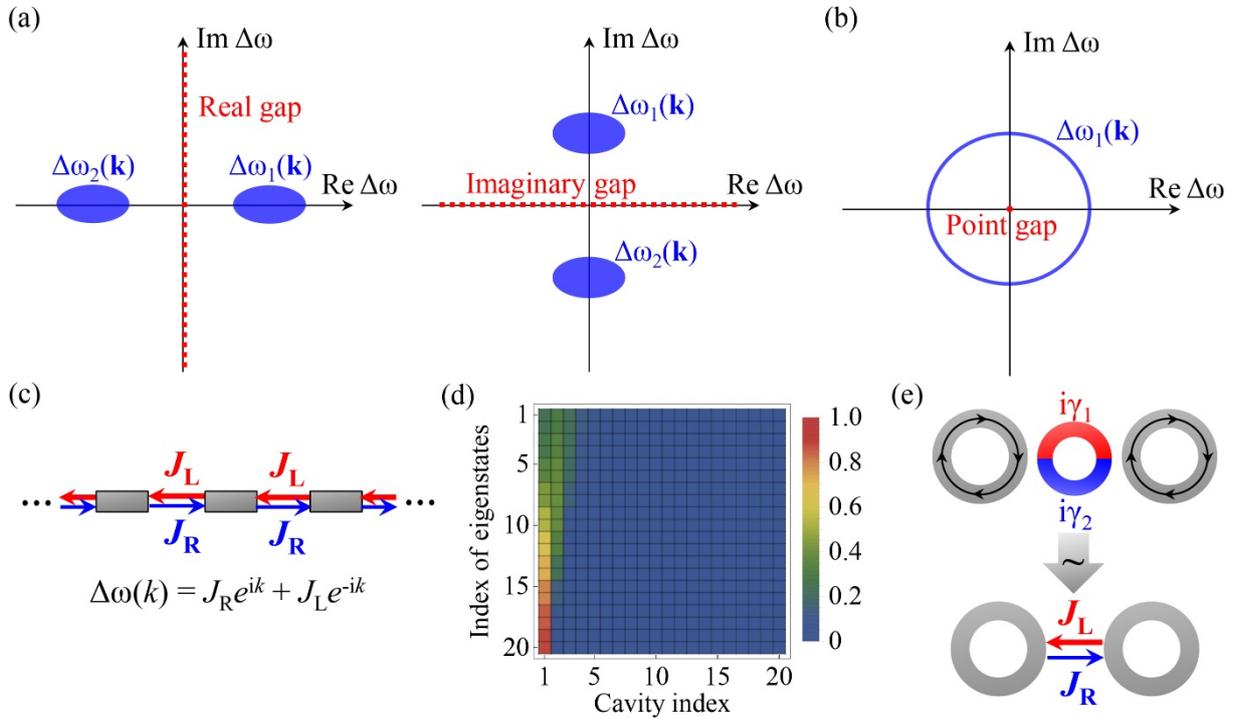

**Fig.11:** (a), (b) Definitions of the complex bandgap. (a) Line gap, where the bulk bands $\Delta\omega_1(\mathbf{k})$ and $\Delta\omega_2(\mathbf{k})$ do not cross a certain line in the complex plane. Left: line gap in terms of $\text{Re}\,\Delta\omega$ (real gap). Right: that in $\text{Im}\,\Delta\omega$ (imaginary gap). (b) Point gap, where the eigenvalues do not touch a certain (symmetry-protected) point in the plane. (d) Fundamental one-dimensional lattice with asymmetric hopping ($J_L > J_R$ here), which exhibits a point gap [129]. (e) Normalized intensity distributions of all the right eigenvectors for a system of 20 cavities with the asymmetric couplings. $J_L = 3$, $J_R = 1$. All the eigenmodes localize around the left edge due to $J_L > J_R$ (non-Hermitian skin effect). (f) Possible realization of asymmetric hopping in a coupled ring cavity system [143,144]. The ancillary ring with distributed gain and loss (centre) induces imaginary gauge fields (effective directional amplification and attenuation) for a supermode with a specific chirality.

# 7 Summary

In this paper, we have reviewed recent progress in active topological photonics. We have particularly discussed topological lasers and non-Hermitian topological physics enabled by gain and loss. We reviewed that various combinations of topological photonic modes with semiconductor gain have been experimentally implemented to realize different types of topological lasers, some properties of which are topologically protected. These prototypical devices point directions for the development of novel lasers with advanced features, such as robust operation under harsh environmental conditions as well as in tightly-limited spaces, unidirectional light output immune to back-reflection noise, and electron-spin-controllable lasing. The exciting physics in topological photonic systems with gain and loss will arguably result in fascinating ideas for further advancing photonic devices. Such non-Hermitian topological photonic systems also offer abundant opportunities to explore novel physics that may be difficult to access using conventional experimental tools in condense matter physics. The field of active topological photonics has just begun and will rapidly grow in the realm of both practical photonic device engineering and fundamental physics.

# Acknowledgements

The authors thank K. Kawabata, The University of Tokyo, for fruitful discussions and M. Parto, D. N. Christodoulides and M. Khajavikhan for providing a schematic illustration. Y.O., Y.A. and S.I. thank MEXT KAKENHI Grant Number JP15H05700, JP15H05868 and 17H06138, and New Energy and Industrial Technology Development Organization (NEDO). Y.O., T.O. and S.I. thank JST CREST Grant Number JPMJCR19T1. K. T. and M. N. acknowledge JST CREST under Grant Number JPMJCR15N4. T.O. is supported by JSPS KAKENHI Grant Number JP18H05857, JST PRESTO Grant Number JPMJPR19L2, RIKEN Incentive Research Project, and the Interdisciplinary Theoretical and Mathematical Sciences Program (iTHEMS) at RIKEN. A.A. acknowledges support from the H2020-FETFLAG project PhoQus (820392), the QUANTERA project Interpol (ANR-QUAN-0003-05), the French National Research Agency project Quantum Fluids of Light (ANR-16-CE30-0021), the French government through the Programme Investissement d'Avenir (I-SITE ULNE / ANR-16-IDEX-0004 ULNE) and the Métropole Européenne de Lille (MEL) via the project TFlight. B.K. acknowledges Office of Naval Research Young Investigator Award N00014-17-1-2671 and NSF Career Award ECCS-1554021.